\documentclass[twocolumn]{aastex63}
\usepackage{graphics,epsf}
\usepackage{amsmath}                % American Mathematical Society package
\usepackage{amsfonts}               % American Mathematical Society fonts
\usepackage{amssymb}                % American Mathematical Society symbol
\usepackage{epsfig}                 % EPS figures
\usepackage{appendix}
\usepackage{graphicx}
\usepackage{float}
\usepackage{color}
\usepackage{multirow}
\usepackage{colortbl}
\usepackage[para,online,flushleft]{threeparttable}

\hypersetup{citecolor=blue, % color for \cite{...} links
            linkcolor=red, % color for \ref{...} links
            menucolor=blue, % color for Acrobat menu buttons
            urlcolor=blue}  % color for \url{...} links

\newcommand{\cm}{{~\rm cm}}
\newcommand{\m}{{~\rm m}}
\newcommand{\km}{{~\rm km}}
\newcommand{\s}{{~\rm s}}
\newcommand{\h}{{~\rm h}}

\newcommand{\g}{{~\rm g}}
\newcommand{\G}{{~\rm G}}
\newcommand{\K}{{~\rm K}}

\newcommand{\yr}{{~\rm yr}}
\newcommand{\Myr}{{~\rm Myr}}

\newcommand{\pc}{{~\rm pc}}
\newcommand{\kpc}{{~\rm kpc}}

\def \astrobj#1{#1}
\begin{document}

\title{Jittering jets by negative angular momentum feedback in cooling flows}

%\correspondingauthor{Noam Soker}
%\email{soker@physics.technion.ac.il}

\author{Noam Soker}
\affiliation{Department of Physics, Technion, Haifa, 3200003, Israel;  soker@physics.technion.ac.il}

\begin{abstract}
I apply the jittering jets in cooling flow scenario to explain the perpendicular to each other and almost coeval two pairs of bubbles in the cooling flow galaxy cluster RBS 797, and conclude that the interaction of the jets with the cold dense clumps that feed the supermassive black hole (SMBH) takes place in the zone where the gravitational influence of the SMBH and that of the cluster are about equal. According to the jittering jets in cooling flow scenario jets uplift and entrain cold and dense clumps, impart the clumps velocity perpendicular to the original jets' direction, and `drop' them closer to the jets' axis. 
The angular momentum of these clumps is at a very high angle to the original jets' axis. When these clumps feed the SMBH in the next outburst (jet-launching episode) the new jets' axis might be at a high angle to the axis of the first pair of jets. I apply this scenario to recent observations that show the two perpendicular pairs of bubbles in RBS 797 to have a small age difference of $<10 \Myr$, and conclude that the jets-clumps interaction takes place in a distance of about $\approx 10-100 \pc$ from the SMBH. Interestingly, in this zone the escape velocity from the SMBH is about equal to the sound speed of the intracluster medium (ICM). I mention two other clusters of galaxies and discuss the implications of this finding. 
\end{abstract}
 

% ==========================================================
\section{Introduction} 
\label{sec:intro}
% ==========================================================
% ==========================================================

When the radiative cooling time of the hot, i.e., X-ray emitting, gas in a galaxy, in a group of galaxies, or in a cluster of galaxies, is shorter than the age of the system some gas cools to very low temperatures and flows in. This is a cooling flow. However, the mass cooling rate to low temperatures is much smaller that the hot gas mass divided by the radiative cooling time  (see however \citealt{Fabianetal2022}).  This inequality requires a heating mechanism. Most studies agree that the heating is by active galactic nucleus (AGN) jets that the central supermassive black hole (SMBH) launches, and that the heating and cooling processes take place in a negative feedback cycle (for reviews see, e.g., \citealt{Fabian2012, McNamaraNulsen2012, Soker2016, Werneretal2019}).
 
The jet feedback mechanism (e.g., \citealt{BaumOdea1991}) contains two main parts. In the cooling part the hot gas cools to form cold clumps that fall/stream inward to feed the SMBH via an accretion disk. In the heating part the accretion disk launches jets that heat the gas. If the cooling rate of the hot gas increases, then more cold clumps fall-in to feed the SMBH leading to more energetic jets that in turn increase the heating rate of the hot gas, therefore reducing cooling. This closes the negative feedback cycle. 

The process by which cold clumps carry the accretion flow onto the central SMBH  (rather than Bondi-type accretion of the hot gas) is the \textit{cold feedback mechanism} 
(\citealt{PizzolatoSoker2005, PizzolatoSoker2010, Sharmaetal2012, Gasparietal2013b, Gasparietal2015, Lietal2015, Prasadetal2015, VoitDonahue2015, Voitetal2015Natur}; note that some studies use other terms for this mechanism, like chaotic cold accretion, e.g., \citealt{Gasparietal2013b, McKinleyetal2022}).
There are many papers in recent years that support the cold feedback mechanism
(e.g., a small sample of recent papers, 
\citealt{
Babyketal2018, Gasparietal2018, Jietal2018, Prasadetal2018, Pulidoetal2018, Voit2018Turb, YangLetal2018, Choudhuryetal2019, Ianietal2019, Roseetal2019, Russelletal2019, Sternetal2019, StorchiBergmann2019, Vantyghemetal2019, Voit2019, 
HardcastleCroston2020, Martzetal2020, Prasadetal2020, Eckertetal2021, Maccagnietal2021, Pasinietal2021, Qiuetal2021, Singhetal2021, Olivaresetal2022}). 

Although there is an agreement that jets heat the hot gas, there is no consensus on the mechanism by which the jets and the bubbles they inflate transfer energy to the hot gas. In what follows I will concentrate on the intracluster medium (ICM), but arguments hold also for cooling flows in galaxies. 
In one group of heating processes the jets and jet-inflated bubbles do work on the ICM by (1) driving shock waves that propagate into the ICM (e.g., \citealt{Randalletal2015, Guoetal2018}), and/or (2) exciting sound waves (e.g., \citealt{Fabianetal2006, TangChurazov2018}), and/or (3) generating internal waves in the ICM by buoyantly rising bubbles (e.g., \citealt{Zhangetal2018}), and/or (4) powering turbulence in the ICM (e.g.,  \citealt{DeYoung2010, BanerjeeSharma2014, Gasparietal2014, Zhuravlevaetal2018}), and/or (5) by uplifting gas from inner regions (e.g., \citealt{GendronMarsolaisetal2017,HuskoLacey2022}).
The four first heating processes cause disturbances that expand to very large distance from the jets and bubbles. 

In the second group of processes the heating of the ICM is more local to the surroundings of the jets and the bubbles. These processes are (6) heating by cosmic rays that the jet-inflated bubbles accelerate and that stream into the ICM and heat it (e.g. \citealt{FujitaOhira2013, Pfrommer2013, Ehlertetal2018, Ruszkowskietal2018, KempskiQuataert2020}) and (7) heating by mixing hot bubble gas (thermal gas and/or cosmic rays) with the ICM (e.g., \citealt{BruggenKaiser2002, Bruggenetal2009, GilkisSoker2012, HillelSoker2014, YangReynolds2016b}). Vortexes that the inflation process of the bubbles forms are responsible for this mixing (e.g., \citealt{GilkisSoker2012, HillelSoker2014}). In \cite{Soker2019CR} I argued that even if the content of the bubble is cosmic rays, the main process that transfers the energy of the cosmic rays to the ICM is mixing and not streaming. 

I take the view that the most efficient heating mechanism is heating by mixing  although many do not accept it, e.g., \cite{SokerHillel2021}  and \cite{Brienzaetal2021} for the most recent dispute.

The most efficient heating of the ICM by mixing takes place when the jets' axes change their directions (e.g., \citealt{Soker2018Jit, Cieloetal2018}). For that, in \cite{Soker2018Jit} I proposed that the feedback cycle includes a process where the jets' axes change direction over time, the jittering jets in cooling flow scenario. Namely, in addition to influencing the energy and mass of the feedback cycle, the jets influence also the angular momentum of the accreted cold clumps.
I will elaborate on this in section \ref{sec:AngularMomentum}, and use new results by \cite{Ubertosietal2021} to refine this scenario for the case of the the cooling flow cluster \astrobj{RBS 797} (section \ref{sec:InteractionZone}). 

\cite{Babuletal2013} previously studies the process by which the central SMBH accretes mass in episodes with stochastic variations of the angular momentum axis. This in turn changes the spin of the SMBH, and hence the direction of the jets' axis. Their motivation was to account for a uniform heating of the ICM with jets. \cite{Babuletal2013} further mention some specific cooling flow clusters where observations show jet reorientation and study the way by which the accretion gas changes the spin of the SMBH. The feeding process of the SMBH in the present study is similar to their picture, but it is not stochastic as I consider that the jets influence the direction of the jets in the next jet-launching episode.  

In a recent thorough and enlightening study of the cooling flow cluster \astrobj{RBS 797}, \cite{Ubertosietal2021} find two pairs of bubbles with the two axes of the pairs almost perpendicular to each other. Moreover, the ages of the two pairs is within $\Delta t_{\rm p} \la 10 \Myr$ from each other.
They suggest that the almost coeval outbursts (jet-launching episodes) that form two perpendicular pairs of bubbles resulted from two black holes (for a theoretical study see, e.g., \citealt{LiuFK2004}) or from a reorientation event of the jets' axis (see \citealt{Gittietal2013} for suggesting these two possibilities for \astrobj{RBS 797}). Each SMBH of the binary launched one pair of jets. \cite{Ubertosietal2021} do not explain why the two pairs of jets are perpendicular to each other. 
I take their results to support the jittering jets in cooling flow scenario  flow that I suggested \citep{Soker2018Jit},  but note that it might be just a stochastic accretion as \cite{Babuletal2013} suggested.  However, in the 2018 Research Note I did not specify the zone in the cluster cooling flow where the jets interact with the clumps. The analysis of \cite{Ubertosietal2021} allows me to locate the zone where the interaction takes place (section \ref{sec:InteractionZone}). 

Previous observations already indicated the reorientation of active galactic nucleus (AGN) jets' axes on a time scale of few~$\Myr$  (e.g., \citealt{DennettThorpeetal2002, LalRao2007}). \cite{DennettThorpeetal2002} attribute the reorientation of the jets' axis to the accretion disk interaction with the SMBH spin as suggested by \cite{NatarajanPringle1998}. This process needs no binary SMBH companion. 

Here I rather explore a third possibility where one pair of jets influences the orientation of the axis of the next pair of jets (section \ref{sec:AngularMomentum}). This process is a result of the jet-feedback mechanism that takes place in cooling flows in galaxies, in groups of galaxies, and in clusters of galaxies, and from the cold-feedback mechanism where cold clumps feed the AGN. 

A very important note is in place here. The jets that I consider in this study are wide-slow-massive collimated outflows, as in the simulations of, e.g., \cite{HillelSoker2014} for cooling flows and of \cite{Fujita2022} for the Fermi bubbles in our Galaxy, and not very narrow relativistic jet that is observed in radio emission. The important point is that such outflows are inferred from observations (e.g. \citealt{DeKooletal2002, Milleretal2020, Choietal2022}). 
\cite{Milleretal2020}, for example, infer an outflow at a velocity of $3150 \km \s^{-1}$ at $9 \pc$ from the AGN of a Seyfert 1 Galaxy. 
\cite{Choietal2022} study  50 low redshift broad absorption-line quasars and find velocities in the range of $\simeq 100 - 10^4 \km \s^{-1}$ at $< 10 \pc$ from the central engine. These and other studies find that $\simeq 20 \%$ of the systems have strong enough outflow to operate a feedback cycle. From that they deduce that the covering solid angle of the outflow is $\Omega \simeq 0.2 (4 \pi)$. This corresponds to jets with a half opening angle of $37^\circ$, i.e., wide jets. 

In section \ref{sec:InteractionZone} I point to the interesting finding that the jets-clumps interaction zone in \astrobj{RBS 797} takes place in the region where the gravitational role of the SMBH and of the cluster are about the same.  I also comment to this equality for two other clusters of galaxies. My crude calculations should be confirmed by future three-dimensional hydrodynamical simulations. I summarise in section \ref{sec:Summary}.

% ==========================================================
\section{Jittering jets in cooling flows} 
\label{sec:AngularMomentum}
% ==========================================================

% ==========================
\subsection{Uplifting dense clumps and feeding the SMBH} 
\label{subsec:Uplifting}
% ==========================

Consider an accretion episode where the central SMBH accretes cold clumps from the ICM, i.e., the cold feedback mechanism. I base the mechanism I consider on the presence of clumps that are cooler than the ICM. Therefore, there must be some clumps in the ICM. In some clusters they might be below detection limit. In those clusters I predict that very deep observations will reveal such clumps close to the center. 

I consider in this study the case where the angular momentum of the ICM is negligible. Therefore, the clumps that are formed by thermal instabilities in the ICM are born with very low angular momentum (but not zero, as they do form an accretion disk around the SMBH).   
  
I define equatorial plane-1 to be the plane where the clumps form the first accretion disk, disk-1, and polar-1 the polar directions along which the disk-1 launches the first pair of jets (first outburst). I follow and expand the arguments from \cite{Soker2018Jit}. The jets and the bubbles they inflate uplift dense clumps by pushing the gas out (e.g., \citealt{Popeetal2010, HillelSoker2018}) or by entraining the clumps (e.g., \citealt{Revazetal2008, Popeetal2010, HillelSoker2014}).  Lifting gas can stimulate AGN feedback (e.g., \citealt{Prasadetal2015, McNamaraetal2016}). 

\begin{figure}
\centering
%%% trim={<left> <lower> <right> <upper>}
%%% For using the .pdf file:
\hskip -1.30 cm
\includegraphics[trim= 3.2cm 15.5cm 2cm 3.1cm,clip=true,width=0.92\textwidth]{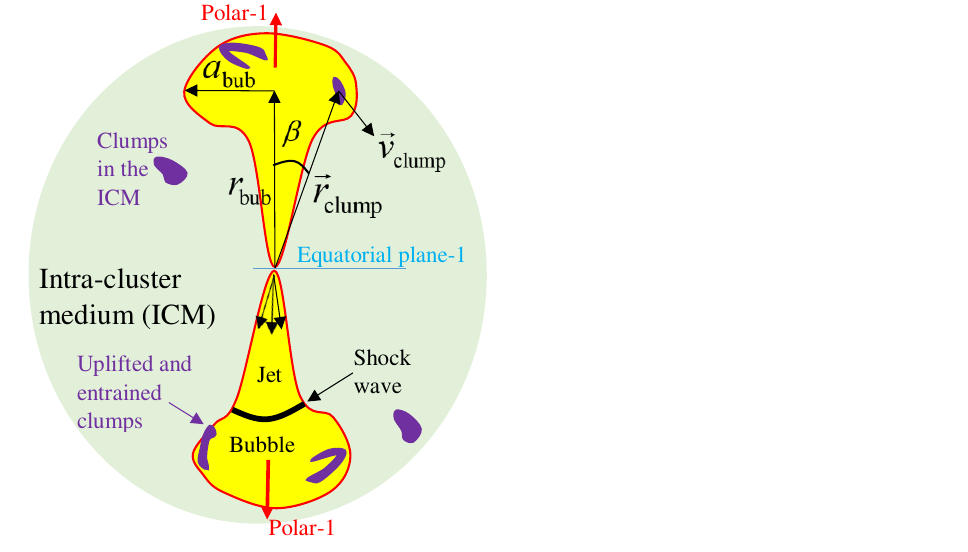}
%%%% For using the .eps file:
%%% \includegraphics[width=0.51\textwidth]{jitteringCFsfig1.eps}
%%% \vskip 0.50 cm
\caption{A schematic drawing in a meridional plane of the flow structure where jet-inflated bubbles and jets (yellow) entrain and uplift dense cold gas (in purple). The jets and the bubbles they inflate accelerate clumps mainly in meridional planes. As such, each clump acquires angular momentum at a large angle to the original jets' axis (polar-1). When the SMBH accretes these clumps their angular momentum sums-up to form an accretion disk with an angular momentum axis (polar-2) at a large angle to polar-1. Therefore, the jets of the next outburst (jet launching episode) will have their axis (polar-2) at a large angle to polar-1. }
\label{fig:flow}
\end{figure}

The vortices that the jets and the jet-inflated bubbles excite entrain the clumps, break them, and drag them. The vortices are excited mainly because of the shear between the jet and the ICM. This implies that the main velocity components of the vortices are in planes that contain the jets' axis and are perpendicular to equatorial plane-1, i.e., the vortices are in a meridional planes. Therefore, each dense and cold clump will have its main velocity components after interacting with the jet in the meridional plane that contains the clump and the jet's axis. I present the schematic flow structure in Fig. \ref{fig:flow}.
%% FFFFFFFFFFFFFFFFFFFFFFFFFFFFFFFFFFFFFFFFFFFFFFFFFF
%% FFFFFFFFFFFFFFFFFFFFFFFFFFFFFFFFFFFFFFFFFFFFFFFFFF

 The clump's specific angular momentum component along the original jet direction (polar-1) is practically zero  
\begin{equation}
\left( \overrightarrow{j}_{\rm clump} \right)_{\rm polar-1}= \left( \overrightarrow{r}_{\rm clump} \times \overrightarrow{v}_{\rm clump} \right)_{\rm polar-1} \simeq 0. 
\label{eq:Jclumps}
\end{equation} 
Moreover, even if a clump has a velocity component perpendicular to the meridional plane its angular momentum will be at a high angle to polar-1 direction because the clump is close to the axis of the first pair of jets (equation 1 in \citealt{Soker2018Jit}).

The conclusion is that because the clumps are close to the jets' axis and because the clumps acquire velocity mainly in a meridional plane, the angular momentum direction of each clump is mainly perpendicular to the angular momentum of the disk that launched the jets (polar-1 direction). When the clumps merge to form the second accretion disk, the combined angular momentum direction, polar-2 direction, will be at a very large angle (up to being perpendicular) to the polar-1 directions. 

Again, this conclusion holds for cases where the specific angular momentum (per unit mass) of the ICM is much smaller than what the jets deposit to the clumps (section \ref{subsec:Conditions}). Note that although the clumps acquire angular momentum at a large angle to the polar-1 direction, the clumps have different angular momentum directions in the plane perpendicular to the polar-1 direction. Namely, the scenario I study here requires that the specific angular momentum of the ICM is smaller than the average specific angular momentum of all accreted clumps. Oppositely, in cases where the ICM does have a large specific angular momentum all clumps are born more or less with the same angular momentum direction as that of the ICM. The angular momentum axis of the accretion disk in the next accretion episode will be the axis of the ICM angular momentum, and so is the axis of the jets this disk launches. In those cases different jet-launching episodes (outbursts) will share more or less the same axis. 

% ==========================
\subsection{The jets direction} 
\label{subsec:JetsDirection}
% ==========================

\cite{Babuletal2013} take the jets' axis to be along the SMBH spin axis. For that they require the accreted mass to change the angular momentum axis of the SMBH, which in turn implies that the SMBH should have a small amount of angular momentum. I do not refer here to any specific jet-launching mechanism or to the question of whether the powering of the jets comes from the SMBH spin or from the gravitational energy that the accreted mass releases. I rather refer to observations. 
In any case, in a new paper \cite{Gottliebetal2022} demonstrate with 3D general-relativity magnetohydrodynamic simulations that the jets are along the angular momentum axis of the accretion disk more than they are along the black hole angular momentum axis. 

Observations show several cases of cooling flow clusters where the axis of the jets changes direction from one jet-launching episode to the next (see discussion by \citealt{Babuletal2013}). In the present study I consider each jet-launching episode to results from a different mass accretion episode.  I emphasize here the case of \astrobj{MS0735+7421} that is the most energetic radio active AGN known (e.g., \citealt{Vantyghemetal2014}), with a central SMBH mass of $\ga 1.5 \times 10^{10} M_\odot$ (e.g., \citealt{Dulloetal2021}). The jets' axis in this cooling flow cluster changes its direction by a large angle of $\approx 40-50^\circ$ (e.g., \citealt{Beginetal2022}). 

  The question of whether the jets' axis is along the SMBH spin axis or along the angular momentum axis of the accreted mass is related to the question of whether these jets are powered by accretion of gas or by the SMBH rotational energy. \cite{McNamaraetal2011} addressed the question of the powering of the AGN activity in cooling flow clusters and could not reach a clear conclusion. Namely, both powering processes might take place in different jet-launching episodes. 
In the case of the cooling flow cluster \astrobj{MS0735+7421} with its hyper-massive BH there are claims for either powering by the SMBH spin (e.g., \citealt{McNamaraetal2009}), or by the gravitational energy that the accreted mass releases (e.g., \citealt{SternbergSoker2009}). In either cases, the change in the direction of the jets' axis in \astrobj{MS0735+7421} must come from the accreted mass  (unless there is a binary SMBH system; see section \ref{subsec:MS0735.6+7421}.)  And if a change in the jets' axis can take place in this most energetic radio active AGN known with its hyper-massive BH, then probably it might take place in other cases.    
  
The point here is that it does not matter for the goals of the present study whether the jets are along the SMBH spin axis or along the accreted mass angular momentum axis. The important point is that a change in jets' axis take place in many cooling flow clusters, and that it results from a change in the direction of the angular momentum axis of the accreted mass. 

% ==========================
\subsection{Conditions for jittering jets} 
\label{subsec:Conditions}
% ========================== 

 First I note that not in all cases repeated outbursts are at large angles to each other. In \astrobj{Hydra A} (e.g., \citealt{Wiseetal2006}) the two inner pairs are along the same line, while the outer pair is tilted by a small angle with respect to the axis of the inner two bubbles. The axis of the inner pair of X-ray bubbles in \astrobj{NGC 5813} is tilted by a small angle with respect to the axis of the outer bubbles \citep{Randalletal2015}. 
 Therefore, although the consecutive pairs of jets are not always at large angle to each other, in most cases they are not exactly on the same line.  

 Two sources of angular momentum of the accreted gas determine the tilt angle. The first is the general angular momentum of the ICM. After all, the source of the clumps is the ICM. I mark the specific angular momentum of the ICM in the region where the jets interact with the ICM, $R_{\rm int}$,  by $\overrightarrow j_{\rm ICM}$. Since the central galaxy is elliptical, the angular momentum is much smaller that the interaction radius times the sound speed $C_s$, which is about the stellar random velocity. Namely, 
\begin{eqnarray}
\begin{aligned} 
j_{\rm ICM} = f_{\rm ICM} R_{\rm int} {C_{\rm s}} \qquad {\rm for} \qquad f_{\rm ICM}\ll 1 , 
\end{aligned}
\label{eq:jICM}
\end{eqnarray}
where the equality defines $f_{\rm ICM}$.
 
The second source is the average specific angular momentum that the jets and the bubbles they inflate deposit into the clumps. Assuming that there are $N_{\rm c}$ clumps distributed randomly around the bubble and close to the jets' axis, the sum of the angular momenta of all clumps will be mainly perpendicular to the original angular momentum axis of the jets. Crudely, the specific angular momentum of all clumps combined will be 
\begin{eqnarray}
\begin{aligned} 
\overrightarrow j_{\rm jitter} & \approx \frac{1}{\sqrt N_{\rm c}} 
\left( \overrightarrow{j}_{\rm clump} \right)
\simeq \frac{1}{\sqrt N_{\rm c}} f_{\rm jitter} {v}_{\rm clump} R_{\rm INT}  , 
\end{aligned}
\label{eq:Jjiter}
\end{eqnarray}
where $f_{\rm jitter} \simeq 0.5$ takes into account that in most cases $\overrightarrow{r}_{\rm clump}$ and $\overrightarrow{v}_{\rm clump}$ are not perpendicular to each other and that $\overrightarrow j_{\rm jitter}$ is the component perpendicular to $\overrightarrow j_{\rm ICM}$. In the second equality I made use of equation (\ref{eq:Jclumps}) and took ${r}_{\rm clump} \simeq R_{\rm INT}$.

 Consider a case where after a long time of no jet activity the ICM accretion forms an accretion disk that launches jets along the ICM angular momentum axis. With the assumptions above, the axis of the next jet-launching episode will be at an angle $\theta_{\rm tilt}$ with respect to the first jets' axis. This angle is given by
\small
\begin{eqnarray}
\begin{aligned} 
& \tan \theta_{\rm tilt}   \simeq  
\frac{j_{\rm jitter}}{j_{\rm ICM}} \approx 
\frac{1}{\sqrt N_{\rm c}} 
\frac{f_{\rm jitter}}{f_{\rm ICM}} 
\frac{{v}_{\rm clump}}{C_{\rm s}} \\ & = 0.5
\left( \frac{N_{\rm c}}{100} \right)^{-1/2}
\left( \frac{f_{\rm jitter}}{0.5} \right)
\left( \frac{f_{\rm ICM}}{0.01} \right)^{-1}
\left( \frac{{v}_{\rm clump}}{ 0.1 C_{\rm s}} \right). 
\end{aligned}
\label{eq:ThetaTilt}
\end{eqnarray}
\normalsize
In the second equality I substituted plausible values, but the variations might be large. If angular momentum of the ICM is much lower and/or if there are only several large clumps, then the tilt angle is close to 90 degrees, i.e., above 60 degrees. On the other hand, cases with many clumps and/or a larger ICM angular momentum result in a small tilt angle.

 The conclusion from the discussion above and equation (\ref{eq:ThetaTilt}) is that conditions might vary from cluster to cluster, mainly $f_{\rm ICM}$, and from one bubble-inflation episode to another, mainly in the typical velocities of the clumps perpendicular to the original jet axis, about $f_{\rm jitter} {v}_{\rm clump}$, and in the number of clumps $N_{\rm c}$.  This might account for the different tilt angle from one case to another.

% ====================================================
\section{The jets-clumps interaction zone} 
\label{sec:InteractionZone}
% ==========================================================
% =============
\subsection{The cluster \astrobj{RBS 797}}
\label{subsec:RBS 797}
% =============
I here try to estimate the zone where the jets interact with the clumps in the case of \astrobj{RBS 797}. I emphasize again that I take the jet-clump interaction to be as in the numerical simulations that I cite in section \ref{subsec:Uplifting}, e.g., \cite{HillelSoker2014}. In this interaction the clumps are at a temperatures of $\ga 10^4 \K$, namely, at pressure balance and therefore their density is at most few hundreds times the ambient density. For typical densities in cooling flow clusters the number density of the clumps is $< 1000 \cm^{-3}$. For example, the clump of density $\simeq (4-6) \times 10^5 \cm^{-3}$ that \cite{Kinoetal2021} study along the jet of \astrobj{NGC 1275} is too dense for the present mechanism. In the process I refer to here the clumps are partially mixed to the bubbles, and the bubbles drag the clumps out. 

A clump falling from radius $R_{\rm int}$ at about the sound speed reaches the center in a time of 
\begin{eqnarray}
\tau_f \simeq  10^4 
\left( \frac {R_{\rm int}}{10 \pc } \right)
\left( \frac {C_{\rm s}}{1000 \km \s^{-1}} \right)^{-1} \yr,
\label{eq:tauf}
\end{eqnarray}
where I scale the sound speed as in \cite{Ubertosietal2021}.  I assume that a dense clump very rapidly is accelerated to the sound speed, but not beyond it because the dissipation is strong for a supersonic velocity. In any case, equation (\ref{eq:tauf}) is an estimate of the timescale that I use below. 
 
\cite{HillelSoker2014} perform hydrodynamical simulations of jets that interact with dense clumps in the ICM. Although they place the dense clumps at $20 \kpc$ from the center, I can scale their findings to much closer clumps as long as they are not too close to the accretion disk that launches the jets. They find that the jets interact with the dense clumps and entrain them, break them to smaller clumps, stretch the clumps, and drag them up within a timescale that is about an order of magnitude longer than the timescale that equation (\ref{eq:tauf}) gives.
\cite{HillelSoker2018} examine the uplifted cool ICM gas by jet-inflated bubbles. However, they did not examine cold clumps that are the subject of the present study. Therefore, I scale by the results of \cite{HillelSoker2014}. 
The clumps will fall later (a process the simulations did not follow). My crude estimate, an estimate that requires future simulations to verify, is that the clumps will start falling at about few, i.e., $\approx 5$, times the entrainment time. Therefore, the time from the first jet activity to the feeding of the SMBH with clumps that the first jets uplifted/entrained is     
\begin{eqnarray}
& \tau_{\rm feed,2} \approx {\rm few} \times 10 \tau_f \approx 
5 \times 10^5 \left( \frac{f_{\rm F}}{50} \right)
\nonumber 
\\ & \times
\left( \frac {R_{\rm int}}{10 \pc } \right)
\left( \frac {C_{\rm s}}{1000 \km \s^{-1}} \right)^{-1} \yr,
\label{eq:taufeed2}
\end{eqnarray}
where the second equality defines the factor $f_{\rm F} \equiv \tau_{\rm feed,2}/ \tau_f$. 

For RBS-797 the age difference between the two outbursts (of the two pairs of bubbles) is $\Delta t_{\rm p} \la 10 \Myr$ or much shorter even. Therefore, the requirement is $\tau_{\rm feed,2} \simeq \Delta t_{\rm p} \la 10 \Myr$. This implies that in the case of \astrobj{RBS 797} the clumps interact with the jets inside a radius of 
\begin{eqnarray}
R_{\rm int} ({\rm RBS}~797)\lessapprox 10 \pc -100 \pc . 
\label{eq:Rint}
\end{eqnarray}

Another requirement is that the time period between the two outbursts (jet-launching episodes) be longer than the lifetime of the accretion disk that feeds the first pair of jets $\tau_{\rm feed,2} > \tau_{\rm disk}$. The reason is that we do not see the jets to precess or change direction continuously. Namely, the first jet-activity (outburst) ends, and only then the second jet activity starts. Let $R_{\rm disk-1}$ be the radius of the disk that feeds the SMBH in the first outburst (first jet-launching episode). I take the viscosity time to be about 100 times the Keplerian orbital time in the disk   (e.g., \citealt{KashiSoker2010arXiv})  and so the time to deplete the accretion disk from mass is 
\begin{eqnarray}
\tau_{\rm disk} \approx 10^4  
\left( \frac {M_{\rm SMBH}}{10^9 M_\odot} \right)^{-1/2}
\left( \frac {R_{\rm disk-1}}{0.1 \pc} \right)^{3/2}
\yr . 
\label{eq:TauDisk}
\end{eqnarray}
For the scaling above the disk radius is about thousand times the Schwarzschild radius of the SMBH, but the disk might be smaller even.

For the jets to interact with the clumps, the clumps should not be too close to the equatorial plane-1. For that, the clumps should still be at a much larger radius than the radius where they form the accretion disk, i.e., where the centrifugal acceleration is still very weak.
This condition is $R_{\rm int} \gg R_{\rm disk-1}$. 

We therefore derive two conditions on the zone of jets-clumps interaction for the proposed scenario. The first is that the clumps that the jets interact with be at $R_{\rm int} \lessapprox 10-100 \pc$ (equation \ref{eq:Rint}) and the second is that they be at 
\begin{equation}
    R_{\rm int} \gg 0.1 \pc \ga R_{\rm disk-1},
\label{eq:Rdiskmin}    
\end{equation}  
for the above scaling of the SMBH mass (but see below). 
This claim for a compact accretion disk, i.e., radius much smaller than $\simeq 1 \kpc$, is compatible with the study of \cite{Prasadetal2017}. They conclude from their simulations that a compact accretion flow with a short viscous time ought to form.

Another relevant radius is the one where the gravity of the cluster and the black hole are about equal, $R_{\rm C-BH}$, namely, where the escape velocity from the SMBH is about equal to the sound speed $C_{\rm s}$. Its value is   
\begin{eqnarray}
& R_{\rm C-BH} ({\rm RBS}~797)  = \frac {2 G M_{\rm SMBH}} {C^2_{\rm s}} =  8.6   \nonumber 
\left( \frac {M_{\rm SMBH}}{10^9 M_\odot} \right)
\\ & \times
\left( \frac {C_{\rm s}}{1000 \km \s^{-1}} \right)^{-2} \pc.
\label{eq:Rzone}
\end{eqnarray}
I scale the mass of the SMBH as \cite{Cavagnoloetal2011} did, but note that they give the possible range to be  $0.6 \times 10^9 < M_{\rm SMBH} < 7.8 \times 10^9 M_\odot$ for the SMBH in \astrobj{RBS 797}. 
 Namely, the value can be in the range of $5 \pc \la R_{\rm C-BH}({\rm RBS}~797) \la 70 \pc$. 

The likely conclusion is that the jets-clumps interaction takes place in the zone of 
\begin{equation}
    R_{\rm int} ({\rm RBS}~797) \approx R_{\rm C-BH} ({\rm RBS}~797).
\label{eq:RintRb}    
\end{equation} 
The volume inner to this zone should be almost depleted of clumps according the the scenario I propose. The implication is that dense clumps evolve relatively slowly while in the ICM, but when they come under the gravitational influence of the SMBH they fall rapidly inward.

Equation (\ref{eq:RintRb}) is the main result of this study. 

% =============
\subsection{The Perseus cluster}
\label{subsec:Perseus}
% =============
 \cite{Dunnetal2006} give for the time difference between the ghost (very outer) bubbles and the outer bubbles in the Perseus cluster a timescale of $\Delta t_{\rm p}({\rm out}) \simeq 6 \times 10^7 \yr$, while the time difference between the inner bubble-pair and the outer bubble-pair is only $\Delta t_{\rm p}({\rm in}) \simeq 6 \times 10^6 \yr$. From these values and equation (\ref{eq:taufeed2}) I find that equation (\ref{eq:Rint}) holds also for the time difference of the inner bubbles of the Perseus cluster, and can be somewhat larger for the outer bubbles. Over all, and very crudely,  
\begin{eqnarray}
R_{\rm int} ({\rm Perseus})\lessapprox 10 \pc -500 \pc . 
\label{eq:RintPerseus}
\end{eqnarray}

 The problem is the central black hole mass of the cD galaxy \astrobj{NGC 1275} at the center of the Perseus cluster, because different studies give different values, in the range of $M_{\rm SMBH} \simeq 1.4 \times 10^{7} M_\odot$  to  $M_{\rm SMBH} \simeq 10^{9} M_\odot$ (see discussion by \citealt{Sanietal2018} and \citealt{Yeungetal2022}).
In light of this large uncertainty I cannot do better that use the scaling in equation (\ref{eq:Rzone}) to find that approximately 
$1 \pc \la R_{\rm C-BH}({\rm Perseus}) \la 8 \pc$.

 I conclude that the case of Perseus is compatible with equation (\ref{eq:RintRb}), namely, 
\begin{equation}
    R_{\rm int} ({\rm Perseus}) \approx R_{\rm C-BH} ({\rm Perseus}).
\label{eq:RintRbPPerseus}    
\end{equation}

% =======================
\subsection{The MS 0735.6+7421 cluster} 
\label{subsec:MS0735.6+7421}
% ======================
 
The cluster \astrobj{MS 0735.6+7421} is a difficult case to study because the change of direction might result from a binary SMBH system  \citep{PizzolatoSoker2005BinaryBH}. Therefore, it might be that the main effect is not due to the jittering jets mechanism that I study here. Nonetheless, let me examine the parameters for this cluster.

 The time difference between the two jet-activity episodes in \astrobj{MS 0735.6+7421} is $\Delta t_{\rm p}({\rm MS0735}) \simeq 6 \times 10^7 \yr - 1.7 \times 10^8 \yr$ \citep{Vantyghemetal2014, Biavaetal2021}. Scaling equation (\ref{eq:taufeed2}) with $\Delta t_{\rm p}({\rm MS0735}) = 10^8 \yr$ I find $R_{\rm int} ({\rm MS0735}) \lessapprox 1000 \pc \label{eq:MS0735}$.

  The SMBH mass of this cluster is huge, with estimates from $M_{\rm SMBH}({\rm MS0735}) \simeq 5 \times 10^9 M_\odot$ \citep{McNamaraetal2009} to $\ga 1.5 \times 10^{10} M_\odot$ \citep{Dulloetal2021}. 
Equation (\ref{eq:Rzone}) gives 
$R_{\rm C-BH}({\rm MS0735}) \simeq 40 - 120 \pc$.

 I conclude, when considering the large uncertainties and the claim for a binary SMBH system, that the values for the cluster \astrobj{MS 0735.6+7421} are compatible with 
\begin{equation}
    R_{\rm int} ({{\rm MS0735}}) \approx R_{\rm C-BH} ({\rm MS0735}).
\label{eq:RintRbPMS0735}    
\end{equation}

% ==========================================================
\section{Summary} 
\label{sec:Summary}
% ==========================================================

Motivated by the thorough study of \cite{Ubertosietal2021} I applied the general jittering jets in cooling flow scenario \citep{Soker2018Jit} to the cooling flow cluster \astrobj{RBS 797}. 
 This scenario is based on jets that uplift gas that stimulate the AGN feedback (e.g., \citealt{Prasadetal2015, McNamaraetal2016}), and on falling clumps that change the jets' axis by their angular monentum (e.g., \citealt{Babuletal2013}) 
The thorough study of \cite{Ubertosietal2021} allows me to locate the zone in which the jets from the first outburst (first jet-launching episode) interact with the clumps that later feed the SMBH to launch the jets of the second outburst.  

My proposed scenario for the perpendicular and almost coeval two bubble-pairs of \astrobj{RBS 797} is as follows. 
\begin{enumerate}
    \item A group (or few groups) of clumps fall to the center and feed the SMBH of \astrobj{RBS 797} via an accretion disk of size $R_{\rm disk-1} \la 0.1 \pc$ (equation \ref{eq:Rdiskmin}). Not all clumps reach the center together, and while some clumps reach the center and form the accretion disk that launches the first pair of jets, some clumps are still further out. 
    \item At large distances of $r \ga R_{\rm int}$ the group (or few groups) of clumps do not fall from the equatorial-1 direction, but rather from a large angle to this plane (see Fig. \ref{fig:flow}). Therefore, the jets interact with many clumps, i.e., entrain and uplift the clumps. This weakens and terminates the first jet-launching activity (first outburst). Equation (\ref{eq:Rdiskmin}) gives the lower bound on the jets-clumps interaction zone. 
    \item The jets and the bubbles they start to inflate drag and uplift the clumps. The remnants of the clumps falls back and feed the SMBH at a time $\tau_{\rm feed,2}$ as I estimated in equation (\ref{eq:taufeed2}).
    \item The time period of $\Delta t_{\rm p} <10 \Myr$ between the two outbursts of \astrobj{RBS 797} \citep{Ubertosietal2021} implies the same limit on  $\tau_{\rm feed,2}$, which in turn bounds the jets-clumps interaction zone to $R_{\rm int} ({\rm RBS}~797)\lessapprox (10-100) \pc$ (equation \ref{eq:Rint}). This allows also the first jet-launching episode (outburst) to shut down before the second outburst.  
    \item When the entrained/uplifted remnants of the clumps fall to the center and feed the SMBH the second jet-launching episode (second outburst) occurs. The new angular momentum direction is highly inclined to that of the first outburst (equation \ref{eq:Jclumps}). 
\end{enumerate}

This scenario might account for the almost coeval and perpendicular two pairs of bubbles in \astrobj{RBS 797}. It turned out that the jets-clumps interaction zone that I derived from the observations of \astrobj{RBS 797}, under the assumption that the jittering jets in cooling flow scenario explains the perpendicular and almost coeval two pairs of bubbles, is about equal to the region where the gravitational influence of the SMBH and the cluster are about equal (equation \ref{eq:RintRb}). 
 I found that two other cooling flow clusters have properties that are compatible with the same relation (equations \ref{eq:RintRbPPerseus} and \ref{eq:RintRbPMS0735}). 

 In section \ref{subsec:Conditions} I estimated the tilt angle from one jets launching episode to the next due to the jittering jets in cooling flow scenario (equation \ref{eq:ThetaTilt}). However, I note that other effects can cause the jets' axis to change, like a binary SMBH system, as  \cite{PizzolatoSoker2005BinaryBH} suggested for \astrobj{MS 0735.6+7421}.

In section \ref{sec:intro} I listed the different ICM heating processes. The main challenge of the heating by mixing is to heat the ICM in all directions. Jittering of the jets ease this challenge. The jittering jets in cooling flow scenario directly connects the heating by mixing process with the cold feedback mechanism (section \ref{sec:intro}). 

The equality that equations (\ref{eq:RintRb}), (\ref{eq:RintRbPPerseus}) and (\ref{eq:RintRbPMS0735}) represent is the main finding of this study. I speculate that it holds in some other cooling flow clusters, and I predict the concentration of dense-cold clumps (clouds) that feed the AGN in that zone.     

% ===================================================
\section*{Acknowledgments}
% ===================================================
I thanks the referees for useful comments. 
This research was supported by the Pazy Foundation.

%%%%%%%%%%%%%%%%%%%%%%%%%%%
%\textbf{Data availability}
%The data underlying this article will be shared on reasonable request to the corresponding author.  
%%%%%%%%%%%%%%%%%%%%%%%%%%%

\label{lastpage}
\end{document}